\pdfoutput=1

\documentclass[a4paper,one-column]{article}

\usepackage{chemformula}
\usepackage{bm}

\usepackage{floatrow}
\newfloatcommand{capbtabbox}{table}[][]
\usepackage{amsmath}
\usepackage{amsfonts}

\usepackage[separate-uncertainty=true]{siunitx}
\usepackage{subcaption}

\newcommand{\bta}{BT1}
\newcommand{\btb}{BT2}
\newcommand*{\pkg}[1]{\textsc{#1}}
\DeclareMathOperator{\erf}{erf}
\newcommand{\etal}{et\penalty50\ al.}
\newcommand{\bH}{\ensuremath{\beta_{\text{H}}}}
\newcommand{\bL}{\ensuremath{\beta_{\text{L}}}}
\newcommand{\dH}{\ensuremath{\delta_{\text{H}}}}
\newcommand{\dL}{\ensuremath{\delta_{\text{L}}}}
\renewcommand{\vec}[1]{\ensuremath{\boldsymbol{#1}}}

\usepackage{hyperref}

\hypersetup{
    pdftitle={High-energy X-ray phase-contrast CT of an adult human chest phantom},    
    pdfauthor={Jannis Ahlers},     
    pdfsubject={High-energy X-ray phase-contrast CT of an adult human chest phantom},   
    pdfcreator={Jannis Ahlers},   
    pdfproducer={Jannis Ahlers}, 
    pdfkeywords={phase-contrast} {propagation-based} {lung} {X-ray} {lungman}, 
}

\usepackage[capitalise]{cleveref}
\creflabelformat{equation}{#2\textup{#1}#3}

\usepackage[margin={2cm,2cm}]{geometry}
\title{High-energy X-ray phase-contrast CT of an adult human chest phantom}
\author{Jannis N. Ahlers$^{1,*}$, Lorenzo D'Amico$^{1,2}$, Henriette Bast$^{3,4}$, Lucy F. Costello$^{1}$,\\Martin Donnelley$^{5,6}$, Samantha J. Alloo$^{1}$, Stephanie A. Harker$^{1}$,Ying Ying How$^{1}$,\\Michelle K. Croughan$^{1}$, James A. Pollock$^{1}$, Daniel Hausermann$^{7}$,\\Anton Maksimenko$^{7}$, Christopher Hall$^{7}$, Timur E. Gureyev$^{8}$, Yakov I. Nesterets$^{9}$, \\Marcus J. Kitchen$^{1}$, Konstantin M. Pavlov$^{10,1,11}$, and Kaye S. Morgan$^{1}$}
\date{\small{
$^{1}$ School of Physics and Astronomy, Monash University, Clayton, VIC 3800, Australia\\
$^{2}$ Elettra-Sincrotrone Trieste S.C.p.A., Trieste, Italy\\
$^{3}$ Chair of Biomedical Physics, School of Natural Sciences, Technical University of Munich, 85748 Garching, Germany\\
$^{4}$ Munich Institute of Biomedical Engineering, 85748 Garching, Germany\\
$^{5}$ Adelaide Medical School and Robinson Research Institute, University of Adelaide, Adelaide, South Australia, Australia\\
$^{6}$ Respiratory Medicine, Women’s and Children’s Hospital Adelaide, Adelaide, South Australia, Australia\\
$^{7}$ Australian Synchrotron, ANSTO, Clayton VIC 3168, Australia\\
$^{8}$ School of Physics, University of Melbourne, Parkville, Victoria 3010, Australia\\
$^{9}$ Commonwealth Scientific and Industrial Research Organisation, Clayton, VIC 3168, Australia\\
$^{10}$ School of Physical and Chemical Sciences, University of Canterbury, Christchurch 8140, New Zealand\\
$^{11}$ School of Science and Technology, University of New England, Armidale, NSW 2351, Australia\\
$^{*}$ Corresponding author: jannis.ahlers@monash.edu}}

\begin{document}

\maketitle
\begin{abstract}
    \noindent
    Propagation-based phase-contrast X-ray imaging is a promising technique for in~vivo medical imaging, offering lower radiation doses than traditional attenuation-based imaging. 
    Previous studies have focused on X-ray energies below \qty{50}{\keV} for small-animal imaging and mammography. 
    Here, we investigate the feasibility of high-energy propagation-based computed tomography for human adult-scale lung imaging at the Australian Synchrotron's Imaging and Medical Beamline. 
    This facility is uniquely positioned for human lung imaging, offering a large field of view, high X-ray energies, and supporting clinical infrastructure. 
    We imaged an anthropomorphic chest phantom (LungMan) between \qty{50}{\keV} and \qty{80}{\keV} across the range of possible sample-to-detector distances, with a photon-counting and an integrating detector. 
    Strong phase-contrast fringes were observed with the photon-counting detector, even at high X-ray energies and a large pixel size relative to previous work, whereas the integrating detector with lower spatial resolution showed no clear phase effects. 
    Measured X-ray phase-shifting properties of LungMan aligned well with reference soft tissue values, validating the phantom for phase-contrast studies. Imaging quality assessments suggest an optimal configuration at approximately \qty{70}{\keV} and the longest available propagation distance of \qty{7.5}{\m}, indicating potential benefit in positioning the patient in an upstream hutch. This study represents the first step towards clinical adult lung imaging at the Australian Synchrotron.
\end{abstract}
\bigskip


\section{Introduction}

Since its invention in the late 19th century, X-ray imaging has become an indispensable tool in medicine. 
With the development of bright, coherent X-ray sources that reveal X-ray wavefield self-interference, propagation-based phase-sensitive imaging became possible in the 1990s \cite{snigirev:1995:pxpcmchsr, cloetens:1996:posrhxi,wilkins:1996:piuphx}. 
The imaging setup is illustrated in \cref{fig:setup}. 
A spatially coherent X-ray wavefield passes through a sample and the wavefield intensity is measured some distance downstream.
Attenuation and phase changes introduced to the X-ray wavefront by the sample are quantified by the sample's complex refractive index:
\begin{equation}
    n(\vec{r};E) = 1 - \delta(\vec{r};E) + i\beta(\vec{r};E),
\end{equation}
where $\vec{r} = (x,y,z)$ is the position within the sample, and $\delta$ (which is $\ll 1$) and $\beta$ encode X-ray phase changes and attenuation, respectively.
Both $\delta$ and $\beta$ decrease with X-ray energy $E$, but phase contrast decreases more slowly, making it relatively more important than absorption at higher energies \cite{gureyev:2009:rrrpxpcbi}. 
The phase changes lead to bright and dark interference fringes forming at sample edges, enhancing the visibility of these structures. 
A phase-retrieval algorithm is then typically applied to recover quantitative sample information from the image (e.g.\ Paganin~\etal\,(2002)\ \cite{paganin:2002:spaesdiho}). 
Compared to attenuation-based imaging, PBI significantly improves contrast-to-noise, which is especially important for differentiating structures with similar attenuation, such as different soft tissues \cite{nesterets:2014:npxpict,kitchen:2017:cdrftuxpc}.
PBI has been employed for ex~vivo studies of human organs and tissues \cite{horng:2014:cstiuxppcthkccith, romell:2018:sihmppc}, as well as in~vivo studies with small animals \cite{kitchen:2017:cdrftuxpc,dekker:2019:ire1orprb,morgan:2020:mdsxrila}. 
Much of this imaging has been performed at synchrotron X-ray sources, which offer high coherence, high flux, and low divergence.
However, synchrotron beam sizes are often limited to less than \qty{1}{\cm} \cite{albers:2023:hrplicrxdl}, resulting in a limited field of view.
Additional challenges---including patient positioning and regulatory approval---have meant that the development of in~vivo human PBI has so far been limited. 
Most efforts have focused on PBI for breast imaging, leveraging the strong phase-contrast effects observed at typical mammography X-ray energies of approximately \qtyrange{15}{35}{\keV}\cite{oliva:2020:eoebsr,arhatari:2021:xpctstiimbias}.
The first in~vivo human PBI experiment was performed at the SYRMEP beamline, Elettra Sincrotrone Trieste \cite{longo:2016:btsrefi}, imaging the breast.
A second effort toward clinical PBI breast CT has been underway at the Imaging and Medical Beamline (IMBL) of the Australian Synchrotron \cite{gureyev:2018:sxptbci,lewis:2018:csptbcpci}.
The IMBL was designed with both medical and industrial imaging requirements in mind; 
it combines high X-ray energies (\qtyrange{20}{120}{\keV}) and a large field of view (full width at half maximum of \qty{522}{\mm} W by \qty{19}{\mm} H at \qty{80}{\keV})\cite{::bltsasuoc} with extensive infrastructure for animal and human studies \cite{hausermann:2010:imblas}. 
The presence of a clinical program, together with the large field of view and high energies available at IMBL, makes it the ideal facility for in~vivo PBI beyond breast imaging, particularly of larger body parts where higher-energy X-rays are required for sufficient transmission.

Breast cancer is the most prevalent cancer in women; in men, it is lung cancer\cite{bray:2024:gcs2geimw3c1c}. Lung cancer is the leading cause of cancer death in the world \cite{bray:2024:gcs2geimw3c1c}.
Within the last few years, a consensus supporting low-dose CT screening for lung cancer in high-risk patients has emerged \cite{leiter:2023:gblccsft}. 
Due to the presence of air--tissue interfaces, the lungs produce strong X-ray phase contrast \cite{kitchen:2004:osxpcilt}.
The inherent dose reduction offered by PBI\cite{nesterets:2014:npxpict,kitchen:2017:cdrftuxpc} suggests the technique may offer a significant advantage over conventional low-dose CT for lung screening. 
PBI has already been widely employed for small-animal lung imaging \cite{yagi:1999:rximlusrs,kitchen:2004:osxpcilt,hooper:2007:ilallcb,dubsky:2012:sdcttmrlfm,kitchen:2017:cdrftuxpc,gradl:2018:vdpxiucls,dekker:2019:ire1orprb,morgan:2020:mdsxrila}.
A recent study at the IMBL showed acceptable image quality at clinical doses using PBI for infant-scale lung imaging \cite{pollock:2024:lhcilppc}, and found the optimal imaging energy was approximately \qty{50}{\keV}.
At the scale of an adult human, previous work at SYRMEP showed the feasibility of low-dose high-resolution PBI CT using porcine lungs in a phantom casing shaped like a human chest \cite{wagner:2018:splippsplhcp,albers:2023:hrplicrxdl}.
Due to the limitations of the beamline used in those studies, they were restricted to a maximum energy of \qty{40}{\keV}.
Therefore, an optimal configuration needs to be determined by investigating energies higher than \qty{40}{\keV}, where X-ray transmission will be higher and dose lower.

For this first study of adult-scale whole lung imaging at IMBL, we used an anthropomorphic chest phantom (LungMan; Kyoto Kagagu)\cite{gomi:2011:ccdsdtidsrdspncps,xie:2013:savpnl16mcaps}. 
The LungMan phantom was designed to approximate the human body for X-ray projection and CT imaging, mimicking attenuation so that reconstructed CT images show the correct Hounsfield units (HUs). 
It is intended to be used for training and research in traditional attenuation-based X-ray imaging. 
In addition, LungMan has been found to be an appropriate phantom for dosimetry and automatic exposure control calibration, with reasonable equivalence between the tissue-equivalent materials used in LungMan and ICRP~Publication~89 reference materials \cite{icrp:2002:bapdurprvip8,perez:2018:cvtpldacros}. 
However, we are using propagation-based phase contrast, and the equivalence of LungMan's absorption properties ($\beta$) to real humans does not directly imply a similar equivalence of phase-shifting properties ($\delta$). 
Therefore, a key question for this study was whether the phase-shifting properties of the materials used in LungMan are equivalent to human tissue.
The LungMan phantom was imaged at several propagation distances and X-ray energies, up to a maximum of \qty{80}{\keV}. Imaging was performed using two detectors: a traditional integrating detector and a photon-counting detector. 
We compared results from the two different detectors; first, by qualitative comparison of the observed phase contrast, and second, by comparing the imaging quality obtained with each detector. 
Next, we verified that the soft tissue-equivalent material in LungMan has sufficiently similar phase-shifting properties to reference soft tissue values within the energy range used in this experiment. 
Finally, we compared the imaging quality at different X-ray energies and sample-to-detector propagation distances.
These results represent the first step toward clinical propagation-based imaging of the whole adult-scale lung at IMBL.

\section{Methods}

\subsection{Imaging}\label{sec:method-im}

\begin{figure}
    \centering
    \includegraphics[scale=1]{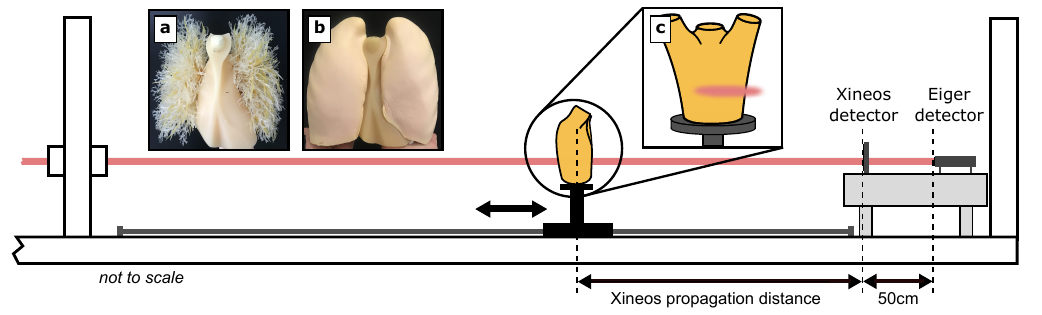}
    {\phantomsubcaption\label{fig:setup-tree}}
    {\phantomsubcaption\label{fig:setup-foam}}
    {\phantomsubcaption\label{fig:setup-beampos}}
    \caption{Experimental setup in hutch~3B of the Imaging and Medical Beamline (IMBL). The CT stage can be moved along the beam path (shown in red). As Eiger is set back by \qty{50}{\cm} on the detector table, the propagation distances for Eiger datasets were always \qty{50}{\cm} more than for Xineos datasets. The inset images show (\subref{fig:setup-tree}) the tree insert, composed of a mediastinum with attached pulmonary vessels, and (\subref{fig:setup-foam}) the foam insert, while the inset diagram (\subref{fig:setup-beampos}) shows the approximate size and position of the beam on LungMan.}
    \label{fig:setup}
\end{figure}

The LungMan phantom is a urethane-based resin male thorax (density: \qty{1.06}{\gram\per\centi\meter\cubed}) with an embedded spine and rib cage composed of epoxy resin mixed with calcium carbonate (density: \qty{1.31}{\gram\per\centi\meter\cubed}). LungMan has a height of approximately \qty{45}{\centi\meter} and a chest circumference of approximately \qty{94}{\centi\meter}. 
The thoracic cavity is empty and can be filled with various inserts. 
In this work, we used two inserts.
The first insert consisted of a mediastinum with attached pulmonary vessels, supported on a solid abdomen/diaphragm block (see \cref{fig:setup-tree}). 
We refer to this as the `tree' insert. 
It is composed of the same urethane-based resin as the main body of the phantom. 
While this insert produces realistic-looking attenuation-based radiographs, it does not contain any materials analogous to the porous lung tissues. 
These tissues are of particular importance in coherent X-ray imaging, as they produce X-ray speckle contrast \cite{yagi:1999:rximlusrs,kitchen:2004:osxpcilt} and X-ray dark-field signals \cite{kitchen:2020:eqmradu2pcxi,willer:2021:xdcidqepcopddas,frank:2021:dfcdcr, gustschin:2022:diccsppbfr}. 
Therefore, a second insert consisting of a resin-based mediastinum with two urethane foam lungs was also imaged (see \cref{fig:setup-foam}).
The urethane foam mirrors the porous structure of lung tissue---we refer to this as the `foam' insert.
Both inserts were used in the comparison of the two imaging detectors, while we focused on the `tree' insert for energy and distance optimisation (see \cref{sec:detector-insert} for more details).

Imaging was performed at the Imaging and Medical Beamline (IMBL) of the Australian Synchrotron, split across two different experiments in August~2022 (referred to as \bta) and July~2024 (referred to as \btb). 
The beamline consists of a superconducting multi-pole wiggler insertion device, which was set to \qty{3}{\tesla} field strength, together with a bent double Laue crystal monochromator providing an energy resolution of $\Delta E/E \approx 10^{-3}$ at the chosen energy.
Imaging was conducted in hutch~3B with the detector table placed against the downstream wall, approximately \qty{145}{\meter} from the source.
The IMBL has two detectors with a wide enough field of view for human adult-scale lung imaging, which we compared for their suitability for lung imaging. 
The first detector was a Teledyne-Dalsa Xineos 3030HR (hereinafter ``Xineos''), an integrating flat-panel detector with \qty{99}{\um} pixel size consisting of a CsI:Tl scintillator with an amorphous silicon photodiode array with an area of $\qty{296}{\mm} \times \qty{296}{\mm}$. 
The second detector was an EIGER2~CdTe~3M\nobreakdashes-W (DECTRIS~AG, Switzerland) (hereinafter ``Eiger''), a hybrid-pixel photon-counting detector with \qty{75}{\um} pixels and an area of $\qty{233}{\mm} \times \qty{80}{\mm}$.
We used an energy threshold with Eiger to ignore any detected photons below the set energy. At \bta, the threshold was set to \qty{25}{\keV} for all imaging. 
However, ideally, the threshold should be set to half the imaging energy to avoid any loss in spatial resolution due to charge sharing. 
At \btb, the threshold was always set to half the imaging energy.
With the goal of optimising the imaging setup for human adult-scale lung imaging, the sample was imaged at combinations of four X-ray energies and four propagation distances.
The energies were \qtylist{50;60;70;80}{\keV}. 
The four propagation distances differed slightly between the two detectors.    
A diagram depicting the positioning of the phantom relative to the two detectors is shown in \cref{fig:setup}.
To adjust the propagation distance, the sample stage was moved along the beam axis while the detector table was kept in place and positioned against the upstream wall of the hutch.
Measured from the centre of the sample stage to the front face of the Xineos detector, the four distances were: \qty{0.25}{\m}, the shortest distance that still allowed the phantom to freely rotate without hitting the detector; \qty{3}{\m}; \qty{5}{\m}; and \qty{7}{\m}, the longest distance possible within hutch~3B.   
As the front face of the Eiger detector is positioned \qty{0.5}{\m} downstream relative to the Xineos detector, the four propagation distances for the Eiger measurements were \qtylist{0.75;3.5;5.5;7.5}{\m}. 
Each CT scan was taken over \qty{360}{\degree}, with an offset centre of rotation in order to capture the full width of the lung. 
At \bta, 7200 projections were collected (step size of \qty{0.05}{\degree}), while at the \btb\ 3600 were collected (step size of \qty{0.1}{\degree}).
Images with Xineos were taken at an exposure time of \qty{33}{\ms}, while those with Eiger were taken with \qty{100}{\ms}.
Note that this meant that each projection taken with Eiger exposed the sample to three times more dose than a projection taken with Xineos. 
For each scan with Xineos, 100 flat-field and dark-current images and averaged.   
As Eiger is a photon-counting detector, only the flat-field images were collected. 

\subsubsection{Image processing}

Before CT reconstruction, a number of pre-processing steps were taken. 
Images were corrected for the dark current (for the Xineos detector only) and the flat field (for both detectors). 
Bad pixels and thin gaps were inpainted using nearest-neighbour interpolation, while larger inter-module gaps in the Eiger images were filled using Navier-Stokes inpainting \cite{bertalmio:2001:nfdivi}.
Combined ring removal was performed on the resulting sinograms \cite{vo:2018:steraxm}. 
Projections separated by 180 degrees were stitched together; first the overlap was manually measured, then one of the projections was flipped horizontally stitched with the other using the calculated overlap.
The stitching did not use any averaging, the two projections were simply sliced and joined at the centre of the overlap.  
The merged projections were then phase retrieved using the Paganin algorithm\cite{paganin:2002:spaesdiho}:
\begin{equation}
    I_\text{phase-retrieved} = \mathfrak{F}^{-1} \left( \frac{\mathfrak{F}[I]/I_\text{in}}{\frac{\Delta\gamma}{2k}|\vec{\xi}|^2 + 1} \right),
\end{equation}
where $\Delta$ is the propagation distance from sample to detector, $k$ is the wavenumber, $\vec{\xi} = (\xi_x, \xi_y)$ are the Fourier-space coordinates, and $\gamma$ is the phase-retrieval parameter. This $\gamma$ parameter was set as the ratio $\delta/\beta$, where $\delta$ and $\beta$ were obtained from reference values for soft tissue (NIST162) \cite{schoonjans:2011:xlxird} at each energy (see \cref{sec:phase-results} for an explanation of the choice of these values). 
Finally, CT reconstruction was performed using filtered back-projection \cite{brun:2017:stpguiccrw}.

\subsection{Dosimetry}\label{sec:dose-method}

Dosimetry was based on a two-step process: first, detector calibration to enable (retrospective) flat-field to incident air kerma conversion, and then Monte Carlo simulations on a digitised and segmented model of LungMan to calculate incident air kerma to mean absorbed dose (MAD) conversion coefficients.

For each detector and monochromator energy used in the experiment, the air kerma rate at the rotation stage position was measured using a thimble ionisation chamber (30010 Farmer chamber, PTW). 
As the unexpanded beam at IMBL rolls off rapidly vertically but is relatively flat horizontally, the ion chamber was oriented horizontally, with the centre of the chamber positioned in the centre of the beam. 
Electrometer readings were integrated for \qty{10}{\second}. 
The conversion from charge to air kerma included corrections for temperature, pressure, and electrometer calibration. 
For each measurement, a flat-field image of the beam without the ion chamber was collected at the same exposure time. 
The flat-field counts were corrected for dark current, and the mean counts in the area of the ion chamber were used to calibrate the flat-field counts to air kerma. 
In the case of the Eiger detector, the calibration procedure was additionally carried out at different threshold values, covering \qtyrange{6}{42}{\keV} in steps of \qty{2}{\keV}. 
When these measurements were later used in calculating the conversion coefficients $C_{\text{ff-KERMA}}$ at a particular threshold level, the measurements were piecewise linearly interpolated across the threshold levels.     

\begin{figure}
    \begin{floatrow}
    \centering
    \ffigbox{
    \includegraphics[scale=1]{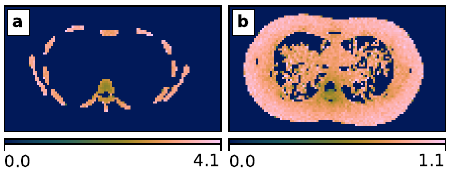}
    {\phantomsubcaption\label{fig:dosedist-bone}}
    {\phantomsubcaption\label{fig:dosedist-st}}
    }{
  \caption{Incident air kerma to \emph{local} absorbed dose conversion coefficients for (\subref{fig:dosedist-bone}) cortical bone and (\subref{fig:dosedist-st}) soft tissue at \qty{70}{\keV}.}\label{fig:dosedist}%
}
\capbtabbox{%
\centering
  \begin{tabular}{ccc}
\hline
Energy (keV) & Cortical bone (CB) & Soft tissue (ST) \\
\hline
50 & \num{2.331\pm0.0013} & \num{0.6587\pm0.00018} \\
60 & \num{2.698\pm0.0014} & \num{0.7621\pm0.00030} \\
70 & \num{2.662\pm0.0025} & \num{0.8147\pm0.00024} \\
80 & \num{2.4375\pm0.00090} & \num{0.8348\pm0.00024} \\
\hline
\end{tabular}
}{%
  \caption{Incident air kerma to mean absorbed dose (MAD) conversion coefficients ($C_\text{CB}$ and $C_\text{ST}$) from Monte Carlo simulation.}
  \label{tab:MCconvcoeff}
}
    \end{floatrow}
\end{figure}

Due to the rapid vertical roll-off of the beam at IMBL, LungMan was exposed to very different levels of incident air kerma within different axial slices of each CT reconstruction volume.
For each reconstruction, the respective averaged flat-field image was averaged horizontally, giving the mean flat-field value $\overline{\text{ff}}$, expressed as counts per pixel, for each slice. 
This was converted to air kerma via $C_{\text{ff-KERMA}}$, correcting for attenuation in air (via object-to-detector distance, $\Delta$) and magnification $M$, and multiplying by the number of projections $N_{\text{proj}}$:
\begin{equation}\label{eq:fftokerma}
   \text{Kerma} = \frac{\overline{\text{ff}} - \overline{\text{df}}}{C_{\text{ff-KERMA}}} \times \frac{1}{e^{-\mu_{\text{Air}} \Delta}} \times M^2 \times N_\text{proj} \times \frac{1}{2}.
\end{equation}
The correction of $1/2$ was included as only (just over) half of LungMan was illuminated in each projection.

To convert the incident air kerma to a mean absorbed dose (MAD), a Monte~Carlo simulation was carried out. 
A CT scan of the whole torso of LungMan was collected at \bta\ using Xineos at \qty{70}{\keV}, with the propagation distance minimised to reduce phase contrast.
The reconstructed CT stack was $4\times$ binned from a voxel size of \qty{99}{\um} to approximately \qty{0.4}{\mm}. 
The binned volume was segmented using \pkg{3D Slicer}\cite{fedorov:2012:3sicpqin} into four regions: the air inside and outside LungMan (both were assigned as air in the Monte Carlo simulation); the urethane-based resin, which was designated as soft tissue (ST, ICRU 44); and the calcium carbonate epoxy resin, which was designated as cortical bone (CB, ICRU 44). 
Assuming a uniform illumination and \qty{360}{\degree} scans, the Monte Carlo simulations produced incident air kerma to mean absorbed dose conversion coefficients for soft tissue ($C_\text{ST}$) and bone ($C_\text{CB}$), listed in \cref{tab:MCconvcoeff}.

A complete measure of the stochastic risk arising from the X-ray exposure would require measuring the mean absorbed doses to the various organs in the thorax and then appropriately weighting and summing to calculate the effective dose.
This would be an important step in future experiments in preparation for human imaging (for example, large animal imaging experiments)\cite{lorenzo}.
We decided that the mean absorbed dose to the soft tissue of LungMan $D_\text{ST}$ sufficed as a conservative analogue of the mean absorbed dose to the lung\cite{angel:2009:drorccetcm} for the purposes of optimising the imaging setup (see \cref{sec:iqa-metrics}); local distributions of dose were also output from the Monte Carlo simulation and an example for \qty{70}{\keV} is shown in \cref{fig:dosedist}.

\subsection{Line profiling with \pkg{fileswell}}\label{sec:fileswell}

\begin{figure}
    \centering
    \includegraphics[scale=1]{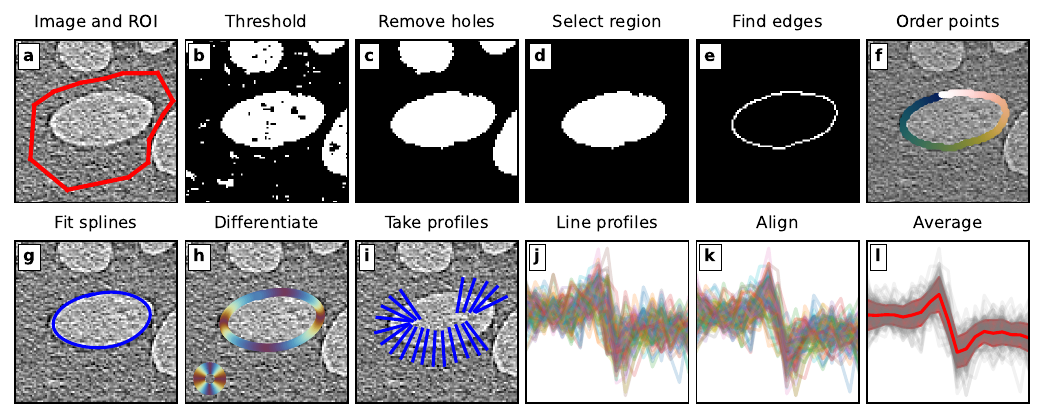}
    {\phantomsubcaption\label{fig:fw-im}}
    {\phantomsubcaption\label{fig:fw-threshold}}
    {\phantomsubcaption\label{fig:fw-holefill}}
    {\phantomsubcaption\label{fig:fw-selectregion}}
    {\phantomsubcaption\label{fig:fw-edges}}
    {\phantomsubcaption\label{fig:fw-order}}
    {\phantomsubcaption\label{fig:fw-spline}}
    {\phantomsubcaption\label{fig:fw-differentiate}}
    {\phantomsubcaption\label{fig:fw-profiles}}
    {\phantomsubcaption\label{fig:fw-lp}}
    {\phantomsubcaption\label{fig:fw-lp_aligned}}
    {\phantomsubcaption\label{fig:fw-lp_mean}}
    \caption{The \pkg{fileswell} algorithm for semi-automated line profiling. (\subref{fig:fw-im}) The input image and region of interest. (\subref{fig:fw-threshold}) Binary thresholding. (\subref{fig:fw-holefill}) Hole filling. (\subref{fig:fw-selectregion}) Selection of the largest region within the field of view. (\subref{fig:fw-edges}) Edge detection using the Canny algorithm. (\subref{fig:fw-order}) Ordering of the edge points using the travelling salesman algorithm. (\subref{fig:fw-spline}) Spline interpolation of the ordered edge points. (\subref{fig:fw-differentiate}) Differentiation of the spline to get the local gradients. (\subref{fig:fw-profiles}) Taking line profiles, that fit within the ROI, across the interface. (\subref{fig:fw-lp}) All line profiles across the interface. (\subref{fig:fw-lp_aligned}) Aligning the line profiles. (\subref{fig:fw-lp_mean}) Averaging to get the mean profile and the standard deviation of the points within that profile.}
    \label{fig:fileswell-method}
\end{figure}

Two aspects of this work required extensive profiling of material interfaces in CT reconstructions: the phase characterisation (see \cref{sec:phase-results}) and the assessment of spatial resolution for optimisation of the imaging setup (see \cref{sec:quality-analysis}). 
A \pkg{python} package, dubbed \pkg{fileswell}, was developed to help automate this process\cite{ahlers:2025:jfirf}. The application takes as input an image, together with a region of interest (ROI) around an interface between two materials, and automatically profiles this interface to give a low-noise averaged line profile. 
For this work, ROIs were selected using the polygon selection tool in \pkg{Fiji}\cite{schindelin:2012:fopba}. 
The line profiling algorithm is outlined in \cref{fig:fileswell-method}. 
It begins by binary thresholding the image within the mask region and removing any small remaining holes and objects. 
The largest contiguous region within the ROI is selected, and a Canny edge detection \cite{canny:1986:caed} is performed to find the interface to be profiled. 
To accurately measure the gradient of the interface at each point, a spline interpolation on the edge points is performed. 
In order to do this we build a graph of the edge points connected to their nearest neighbours and solve the travelling salesman problem to find the optimal ordering. 
The spline interpolation is then performed, and the resulting splines are differentiated to obtain the local gradients. 
Finally, profiles are taken across the interface at each edge point, aligned, and averaged to give a final profile.

\subsection{Phase characterisation}\label{sec:phase-method}

To measure the phase-shifting properties of the soft-tissue equivalent material in the LungMan phantom we used the method of Alloo~\etal\ (2022) \cite{alloo:2022:tpaescumuxppi}. 
Assume that, within the sample, there are two adjacent homogenous materials with complex refractive indices $n_H = 1 - \delta_H+i\beta_H$ and $n_L = 1 - \delta_L+i\beta_L$, where $\bH > \bL$. Alloo~\etal\ model the 1D interface $\beta(x)$ between these two materials (for example, a line profile across the interface in a CT slice) as a modified error function, 
\begin{equation}
    \label{eq:erfmodel}
    \beta(x) = \frac{\bH + \bL}{2} + \frac{\bH - \bL}{2} \erf{\left(\frac{x}{l}\right)},
\end{equation}
where the interface is centred at $x=0$ and $l$ is an interface width. 
This models the interface as a step function that is smoothed, either intrinsically due to intermingling of the two materials at the interface or due to the limitations of the imaging system. 
The step function model applies to a contact image without propagation-based phase contrast.
Consider, in addition, a CT reconstruction of a propagation-based imaging dataset without any phase retrieval applied. 
In this case, the interface between two materials will exhibit a fringe from propagation-based phase contrast. 
We will denote the same line profile in this instance $\beta_\text{PBI}(x)$. 
Following the method in Beltran~\etal\,(2010)\cite{beltran:2010:23xprmousdd} and Alloo~\etal\,(2022)\cite{alloo:2022:tpaescumuxppi}, we apply a transport-of-intensity (TIE) propagator to \cref{eq:erfmodel} to arrive at a model for $\beta_\text{PBI}(x)$:
\begin{equation}\label{eq:alloomodel} 
    \beta_\text{PBI}(x) = \frac{\beta_H + \beta_L}{2} + \frac{\beta_H - \beta_L}{2} \erf{\left( \frac{x}{l} \right)} + \frac{\tau \sqrt{\pi} (\bH - \bL)x}{l^3} \exp{\left( \frac{-x^2}{l^2} \right)},
\end{equation}
where $\Delta$ is the propagation distance from sample to detector, $\lambda$ is the wavelength, and we define $\tau = \frac{\Delta \lambda}{4 \pi} \frac{\dH - \dL}{\bH - \bL}$.
Note that \cref{eq:alloomodel} differs slightly from Eq.\,5 in Alloo~\etal\ (2022) \cite{alloo:2022:tpaescumuxppi}.
Instead of modelling a phase edge in a PBI CT reconstruction without phase retrieval (as in \cref{eq:alloomodel}), Alloo~\etal\ model a phase edge where phase retrieval was previously applied with an incorrect (or non-optimal) retrieval parameter $\gamma'$ (see Eq.\,2 and following text in  Alloo~\etal\ (2022)\cite{alloo:2022:tpaescumuxppi}).
As not doing phase retrieval is equivalent to phase retrieval with $\gamma = 0$, we can recover \cref{eq:alloomodel} from Eq.\,5 in Alloo~\etal\ (2022) by setting $\gamma' = 0$.

The model of $\beta_\text{PBI}(x)$ given in \cref{eq:alloomodel} can be applied to measure the complex refractive index of an unknown sample in a PBI CT dataset \cite{alloo:2022:tpaescumuxppi}. 
First, the CT reconstruction is performed from the non-phase retrieved PBI projections, which still contain propagation-based fringes. A line profile across an interface of the material in the CT is calculated (see \cref{sec:fileswell}), and the model in \cref{eq:alloomodel} is fit to the line profile. The fitting parameters are $\tau$ and the interface width $l$. 
If the materials are sufficiently homogenous, $\beta_H$ and $\beta_L$ can be measured by averaging small regions on either side of the interface and subsequently fixing them during the fitting. 
If one of the two materials has a known refractive index decrement $\delta$, the fitted parameter $\tau$ can then be used to find the other. 
This is simple if the material has an interface to air.

\subsection{Imaging quality analysis}\label{sec:iqa-method}

In order to optimise the imaging quality of setup at IMBL for adult-scale lung imaging, we measured a number of image quality characteristics and combined these into figures of merit.
The characteristics we measured were the signal-to-noise ratio (SNR), contrast, contrast-to-noise ratio (CNR), spatial resolution, and mean absorbed dose. 
We will describe how these were measured in \cref{sec:iqa-measurement}, and then discuss how they were combined into imaging quality metrics in \cref{sec:iqa-metrics}. 

\subsubsection{Measurement of quality characteristics}\label{sec:iqa-measurement}

The measurement of SNR, contrast, and resolution was based on locating regions of interest (ROIs) in the reconstructed CT slices containing homogenous soft tissue with a well-defined boundary to air.
For each CT slice, between one and three interfaces were selected.
An example of selecting such regions in a CT slice is shown in \cref{fig:phaseprop-rois}.
For each ROI, the \pkg{fileswell} algorithm (see \cref{sec:fileswell}) was used to get an average line profile across the interface, as well as mean ($\overline{I}$) and standard deviations ($\text{sd}_I$) of homogenous soft tissue (ST) and air on either side of the interface. 
Then, we calculated the SNR, contrast, and CNR as
\begin{equation}
    \text{CNR} = \text{SNR} \times \text{contrast} = \frac{\overline{I}_\text{ST}}{\text{sd}_{I_{\text{ST}}}} \times \frac{|\overline{I}_\text{ST} - \overline{I}_\text{air}|}{\overline{I}_\text{ST}} = \frac{|\overline{I}_\text{ST} - \overline{I}_\text{air}|}{\text{sd}_{I_{\text{ST}}}}.
\end{equation}
For the measurement of spatial resolution, we assumed the imaging system could be characterised by a Gaussian point spread function.
The spatial resolution was taken as the standard deviation $\sigma$, found by fitting an error function of the form in \cref{eq:erfmodel} to the average line profile across the air--soft~tissue interface.

\subsubsection{Figures of merit}\label{sec:iqa-metrics}

The measured image quality attributes were combined into figures of merit, which could then be directly compared to optimise the setup.
When comparing measurements at different dose levels $D$ (for example, within different slices of a CT stack), the SNR needed to be normalised as $\text{SNR} \propto \sqrt{D}$. 
In addition, the relationship between SNR and spatial resolution $\sigma$ needed to be taken into account. 
A common approach in 2D imaging is to divide the SNR (or CNR) by resolution, giving a figure of merit of $CNR / (\sqrt{D} \times \sigma)$ \cite{reynolds:2022:cdaiqapm}.
Another approach is given by the concept of intrinsic imaging quality $Q$ \cite{gureyev:2016:srsiclis,gureyev:2019:srsrxi}, which is defined as
\begin{equation}
    {Q}^2_\text{sys} = \frac{\text{SNR}^2}{F_\text{in} \times \sigma^{d}},
\end{equation}
where the signal-to-noise ratio (SNR) is measured in a flat field, $d$ is the dimensionality of the image (for example, for voxels in a CT reconstruction $d=3$), the incoming fluence $F_\text{in}$ is the number of photons per $d$-dimensional area, and the spatial resolution $\sigma$ is a width associated with the imaging system's point spread function (PSF).

$Q_\text{sys}$ measures the information gained per photon for a particular imaging system \cite{gureyev:2016:srsiclis}. 
It is the intrinsic imaging quality of an \emph{imaging system}, without consideration of a particular sample.
As such, it does not include the concept of contrast and does not account for biological damage in the form of dose.
By adding these terms, a sample-specific \emph{biomedical} imaging quality can be defined as \cite{gureyevfuture}:
\begin{equation}
    {Q}^2_\text{bio} = {Q}^2_\text{sys} \times C_m^2 \times \frac{\overline{R}_{\text{ab,air}}}{\overline{R}_{\text{ab}}} = \frac{\text{CNR}^2}{\sigma^d} \frac{\overline{R}_{\text{ab,air}}}{\overline{D}_{\text{ab}}}, 
\end{equation}
where $C_m$ is the contrast between two adjacent regions in the image \cite{gureyevfuture}, $\overline{R}_\text{ab} = \frac{\mu}{\rho} E_\text{photon}$ is the mean absorbed dose per (monochromatic) photon in a particular material, and $\overline{D}_{\text{ab}}$ is the mean absorbed dose.  
Like the intrinsic imaging quality, ${Q}_\text{bio}$ is dimensionless.
In the case of three-dimensional (CT) imaging, care must be taken to correctly account for the three-dimensional fluence, as the mean absorbed dose $\overline{D}_{\text{ab}}$ is defined with respect to the two-dimensional fluence \cite{gureyevfuture}:
\begin{equation}\label{eq:Qbio3D}
    {Q}^2_\text{bio,3D} = \frac{\text{CNR}_\text{3D}^2}{\sigma^3} \frac{\overline{R}_{\text{ab,air}}L}{\overline{D}_{\text{ab}}},   
\end{equation}
where the factor $L$ is $\pi / 2$ times the radius of the CT reconstruction cylinder.

Similarly, our measurement of fluence, in the form of the average flat-field counts, gives a two-dimensional fluence, and so the three-dimensional system intrinsic imaging quality is given as
\begin{equation}
    {Q}^2_\text{sys,3D} = \frac{\text{SNR}_\text{3D}^2}{\overline{\text{ff}}},
\end{equation}
where we have taken into account the fact that the 3D incident photon fluence, $F_\text{in,3D}$, is equal to the total number of incident photons during the scan, $N_\text{phot}$, divided by the CT volume, $V_\text{CT}$:  $
    F_\text{in,3D} = {N_\text{phot}}/V_\text{CT}=\overline{\text{ff}}\times N_\text{pix}\times N_\text{proj}/V_\text{CT} \sim \overline{\text{ff}}/\sigma^3$, where $N_\text{pix}$ is the number of pixels in one projection.
Note that $\overline{\text{ff}}$ did not correspond exactly to the number of photons incident per pixel, as the efficiency of Eiger is not 100\%. However, at the energy range of \qtyrange{50}{80}{\keV} used in this experiment the efficiency is approximately constant at 70\% \cite{donath:2023:ehxdasde}.

The end result of this analysis was a table of \num[group-separator = {,}]{10048} measurements, each corresponding to an air--soft-tissue interface within a CT slice.
When averaging measurements taken with a particular combination of detector, energy, and propagation distance, inverse-variance weighting was used, giving the results shown in \cref{sec:quality-analysis}.

\section{Results}

\subsection{Choice of detector and insert}\label{sec:detector-insert}

\begin{figure}
    \centering
    \includegraphics[scale=1]{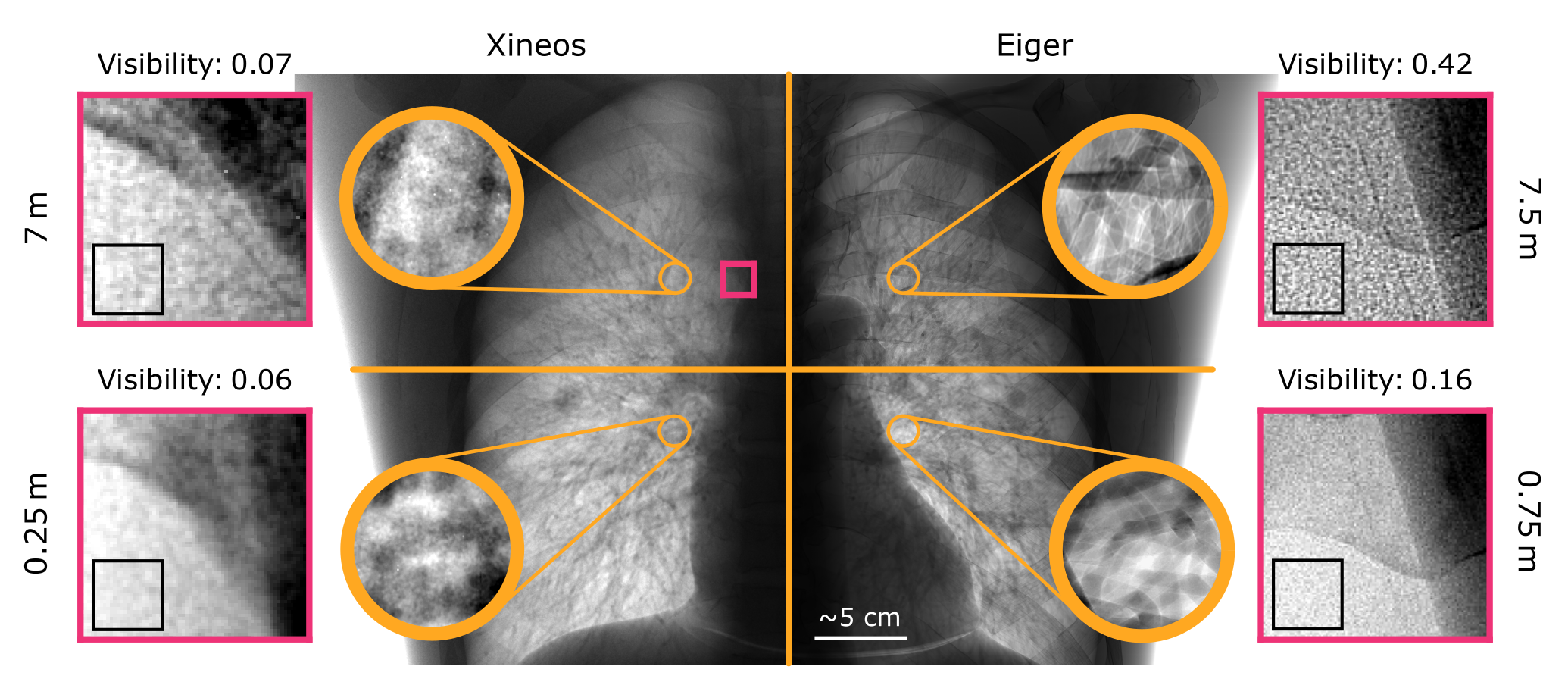}
    \caption{Comparing features in flat-field corrected and stitched projections of LungMan taken with Eiger and Xineos.
    In the central image, Eiger contains the tree insert. It is a composite image, with each quadrant showing a whole-lung stitched projection with a respective detector and propagation distance.
    In addition, we show a small region of interest from images taken of Eiger containing the foam insert (the approximate position of the ROI is shown on the composite image).
    While phase fringes and speckle patterns clearly develop in the Eiger images taken at the longer propagation distance, they cannot be seen with the Xineos detector. 
    Note that the images are stitched from approximately \qtyrange{3}{5}{\cm} tall projections in which the flat-field illumination strongly varied.
    In addition, the Eiger projections were taken with an exposure time three times longer than the Xineos projections.
    Therefore, the dose cannot be directly compared between the different images in this figure; the figure is included to show the qualitative features that were present. 
    The reader is referred to \cref{fig:eigerxineos-Q} for a dose-matched and quantitative comparison of the two detectors. 
    }
    \label{fig:speckles}
\end{figure}

The first step in optimising the imaging setup was choosing between the two X-ray detectors. 
\Cref{fig:speckles} shows projections taken of LungMan with both detectors, at the shortest and longest propagation distances, and with both of the inserts.  
In the centre of the figure is a composite image, showing the full chest of LungMan containing the tree insert. 
Each quadrant of the image was taken with one combination of detector and propagation distance.
In the zoomed insets (outlined in yellow) we can compare the visibility of phase fringes at the edges of the tree structure.
With Xineos, no phase fringes are visible.
On the other hand, with Eiger, fringes can be seen even at the short propagation distance and are particularly pronounced at \qty{7.5}{\m}.
On either side of the central image, we show small regions of interest (outlined in pink) from projections of LungMan containing the foam insert.
The approximate location of the region of interest is marked in the central image. 
While the texture of the foam does not substantially change between the two distances when imaged with Xineos, a strong speckle pattern develops at the longer propagation distance in the Eiger images.
This observation is supported by a measurement of the visibility, defined as $V = (I_{\max}-I_{\min})/(I_{\max}+I_{\min})$, within a small homogenous region in each region of interest.  

\begin{figure}
    \centering
    \includegraphics[scale=1]{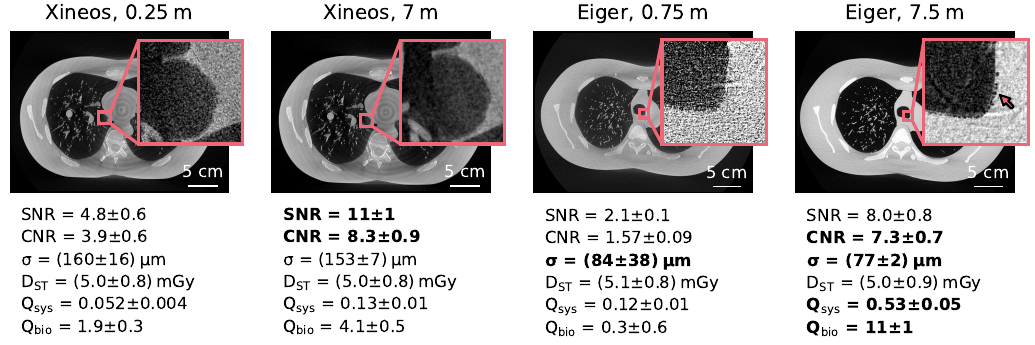}
    \caption{Comparison of phase-retrieved CT slices at the same mean absorbed dose of approximately \qty{5}{\milli\gray}.
    All images were taken at \qty{70}{\keV}.
    See \cref{sec:iqa-method} for definitions of the quality metrics.
    While the signal- and contrast-to-noise ratios are similar for Xineos and Eiger at the longest distances, the significantly better resolution means that imaging qualities are much higher using Eiger at \qty{7.5}{\m} than for any other combination.  
    The zoomed insets highlight the boundary between the mediastinum and the trachea/primary bronchi, which are filled with light foam for structural stability (intended to be transparent in conventional X-ray attenuation imaging). 
    The small air bubbles at the edge of that foam can be clearly seen with Eiger at \qty{7.5}{\m} (see arrow), but are not visible in the other images. 
    All images are shown at the same greyscale.}
    \label{fig:eigerxineos-Q}
\end{figure}

Phase-retrieved and reconstructed CT slices at the same mean absorbed dose of approximately \qty{5}{\milli\gray}, taken with Eiger and Xineos at \qty{70}{\keV} and at the shortest and longest propagation distances, are shown in \cref{fig:eigerxineos-Q}. 
Quality metrics are listed below the images, with the best performance as measured by that metric shown in bold. 
Both the quantitative metrics and qualitative features in the images suggest that Eiger is superior to Xineos for phase-contrast imaging of large samples. Based on these results, we decided to focus on Eiger for the rest of the analysis, and recommend that Eiger be preferred over Xineos for future lung imaging at IMBL.

In choosing a detector, we compared images with both the tree and foam inserts in LungMan. 
While the foam did create speckles (see \cref{fig:speckles}), it is not designed to replicate the fine structure of lung alveoli.
Indeed, several previous studies used LungMan in lung dark-field imaging studies \cite{frank:2021:dfcdcr,andrejewski:2021:r3ixdilfv,gustschin:2022:diccsppbfr,viermetz:2022:dctrhs,bushe:2023:dmcxrmpsdld}, but replaced the insert with custom materials such as neoprene foam \cite{gustschin:2022:diccsppbfr,viermetz:2022:dctrhs} or cotton wool \cite{frank:2021:dfcdcr}. 
In addition, the foam insert lacks the realistic attenuation and heterogeneity provided by the tree insert. 
The pores of the foam were not well resolved in the CT reconstructions.
Finally, the tree insert provided a multitude of clear air--tissue interfaces that could be used in the quality analysis.
As a result, we decided to use the tree insert to evaluate image quality at multiple distances and energies in \cref{sec:quality-analysis}. 

\subsection{Measuring refractive index of LungMan}\label{sec:phase-results}

\begin{figure}
    \centering
    \includegraphics[scale=1]{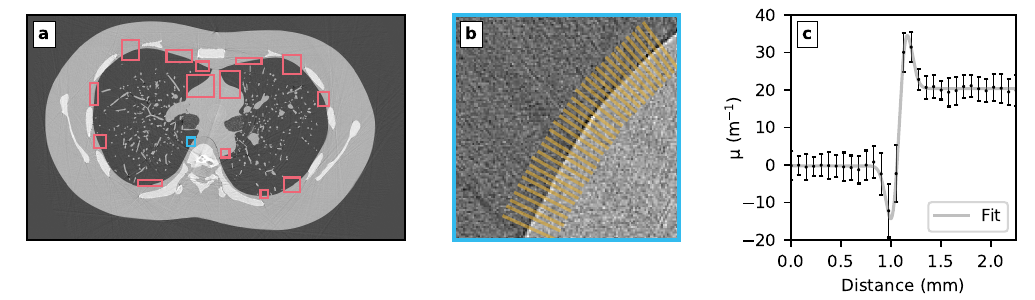}
    {\phantomsubcaption\label{fig:phaseprop-rois}}
    {\phantomsubcaption\label{fig:phaseprop-fileswell}}
    {\phantomsubcaption\label{fig:phaseprop-alloofit}}
    \caption{Measurement of the phase-shifting properties of LungMan. (\subref{fig:phaseprop-rois}) In a central (high dose, low noise) slice of the dataset, a number of air--soft~tissue interfaces are manually selected. (\subref{fig:phaseprop-fileswell}) For each region of interest, the fringe between the two materials is profiled using the \pkg{fileswell} algorithm. (\subref{fig:phaseprop-alloofit}) The resulting fringe profile is fit with a modified error function model (\cref{eq:alloomodel}) to find $\delta$.}
    \label{fig:phase-method}
\end{figure}

\begin{figure}
    \centering
    \includegraphics[scale=1]{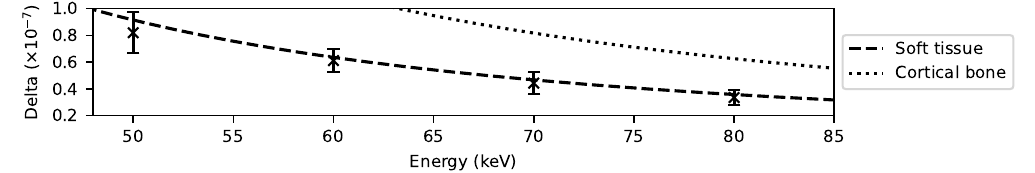}
    \caption{Results of the measurement of delta ($\delta$) values of the soft tissue equivalent material in LungMan at \qtylist[list-units=single]{50;60;70;80}{\keV}. The results show excellent agreement with reference values for soft tissue \cite{internationalcommissiononradiationunitsandmeasurements:1989:tsrdm}. For context, reference values for cortical bone, which differ strongly from the measured values, are also shown. Error bars on the measurements show one standard deviation.}
    \label{fig:phase-results}
\end{figure}

For the characterisation of the X-ray properties of the urethane-based resin that composes LungMan's soft tissues (both the body and the tree insert), we used the CT datasets collected using Eiger at all four energies and at \qtylist{3.5;5.5;7.5}{\meter} propagation distances. 
In each CT dataset, the central slice with the highest flat-field counts/dose was selected for analysis. 
Regions containing homogenous air and tissue were manually selected in each slice.
A representative example containing 15 interfaces is seen in \cref{fig:phaseprop-rois}.  
The line profiles were then fit to the model in \cref{eq:alloomodel} using the Levenberg-Marquardt method \cite{newville:2014:lnlmcp}.
The mean of all resulting delta ($\delta$) values at each energy were taken, and are shown plotted in \cref{fig:phase-results}. 
Comparing with reference values for soft tissue and cortical bone \cite{internationalcommissiononradiationunitsandmeasurements:1989:tsrdm,schoonjans:2011:xlxird} shows that the X-ray phase-shifting properties of the Lungman soft tissue-equivalent material are a good match for soft tissues, and hence Lungman is an appropriate model for studies of X-ray phase contrast imaging.

\subsection{Energy and distance optimisation}\label{sec:quality-analysis}

\begin{figure}
    \centering
    \includegraphics[scale=1]{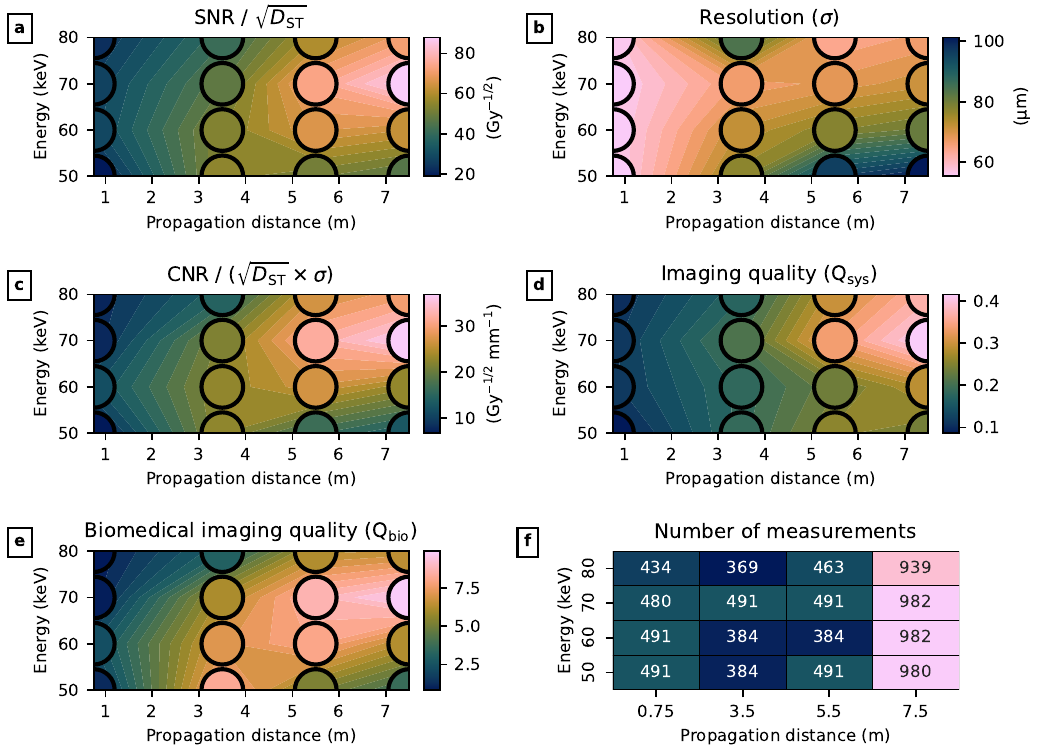}
    {\phantomsubcaption\label{fig:iqaresults-snr}}
    {\phantomsubcaption\label{fig:iqaresults-res}}
    {\phantomsubcaption\label{fig:iqaresults-cnrres}}
    {\phantomsubcaption\label{fig:iqaresults-Qsys}}
    {\phantomsubcaption\label{fig:iqaresults-Qbio}}
    {\phantomsubcaption\label{fig:iqaresults-other}}
    \caption{Results of imaging quality analysis.
    The metrics outlined in \cref{sec:iqa-metrics} are plotted against X-ray energy and propagation distance. The mean value at each combination is shown in the filled circle.  
    As expected, larger propagation distances show increased SNR, but also reduced spatial resolution. 
    Combining fluence-/dose-normalised SNR/CNR and resolution (\crefrange{fig:iqaresults-cnrres}{fig:iqaresults-Qbio}) suggests an optimal setup at \qty{70}{\keV} and \qty{7.5}{\m}.
    \label{fig:iqa-results}}
\end{figure}

After choosing the detector and an appropriate phantom insert, the next step was optimising the X-ray energy and sample-to-detector propagation distance.
Using the Eiger detector, we imaged LungMan with the tree insert at the four energies and distances described in \cref{sec:method-im}, and measured quality characteristics within the reconstructed CT slices as described in \cref{sec:iqa-method}.

As has been noted, the beam roll-off at IMBL meant that different slices in the CT reconstructions were associated with very different mean absorbed doses to the soft tissue. 
These ranged from \qty{1.4\pm0.2}{\milli\gray} (near the edge of the detector, at \qty{80}{\keV}) to \qty{1.1\pm0.1}{\gray} (in the centre of a \qty{50}{\keV} scan). 
Noting that the imaging quality metrics outlined in \cref{sec:iqa-metrics} include a normalisation for dose (or fluence), we plotted the metrics against dose in Supplementary Fig.\,S1.
As these showed a flat dependence, and in order to maximise the statics, we included all measurements in the imaging quality analysis shown here.
Supplementary Section\,S1 includes a re-analysis restricting the measurements to those with a mean absorbed dose below \qty{50}{\milli\gray}.
The outcome in terms of optimal settings is unchanged for this case.

The results are shown in \cref{fig:iqa-results}. 
In \crefrange{fig:iqaresults-snr}{fig:iqaresults-Qbio} we plot the metrics against X-ray energy and propagation distance.  
In these plots, the mean measured values are shown in coloured circles drawn on top of an interpolated contour.
As expected, dose-normalised SNR is maximised when using a long propagation distance. 
By contrast, the spatial resolution of the phase-retrieved CT slices worsens with increasing distance and decreasing X-ray energy, due to an increase in penumbral blur and a broadening of the phase contrast fringes. 
When the SNR/CNR and resolution are combined in \cref{fig:iqaresults-cnrres} and \cref{fig:iqaresults-Qsys}, the quality peaks at \qty{70}{\keV}. 
It also increases with propagation distance, reaching a maximum at the longest distance possible within the limits of hutch~3B at IMBL.

\section{Discussion}

Our results imply that an optimal energy for monochromatic adult-scale thoracic PBI is approximately \qty{70}{\keV}. 
Given the usual rule of thumb that the mean energy of an X-ray tube spectrum is approximately one-third to one-half of the kVp, and that standard chest CT voltages are between \qty{120}{\kilo\volt} and \qty{140}{\kilo\volt} \cite{diederich:2000:reaic}, this result matches well with clinical practice. 
Further work should be done to test if the variability in patient body shapes would justify patient-specific adjustment.

Unlike the results for imaging energy, we could not identify an optimum propagation distance. 
Indeed, previous work on adult-scale lung PBI used a substantially longer propagation distance of \qty{10.5}{\m} \cite{albers:2023:hrplicrxdl}, even with X-rays of much lower energy.  
The main limiting factor with an increasing propagation distance is the associated increase in penumbral blurring due to the finite effective size of the source $\sigma_\text{source}$. 
The penumbral blur width $U$ is given by
$U = \sigma_\text{source}\frac{\Delta}{\text{sod}}$, where $\text{sod}$ is the distance from the source to the object. 
While the effective source size at IMBL varies in the vertical and horizontal directions and is dependent on factors such as the monochromator crystal bend, let us take a conservative upper bound of \qty{1}{\mm} \cite{murrie:2014:fsppxliimbas}. 
With the detector placed at the downstream end of hutch~3B, we would get a penumbral blur width $U = \qty{75}{\um}$ (equal to one pixel on the Eiger detector) at a sample-to-detector distance of approximately \qty{10}{\m}. 
This is longer than can be achieved within hutch 3B and is longer than any of our measurements.     
While a longer propagation distance would not be possible within hutch~3B of the IMBL, there is the option of placing the sample in hutch~3A while keeping the detector in 3B, in order to achieve higher propagation distances. This would require careful testing to ensure that the additional beamline elements between the hutches do not significantly affect the beam quality. While hutch~3A is not currently configured with a sample stage, the idea should be explored for future clinical use.
Note that if a smaller pixel size were used for region-of-interest imaging, the optimal propagation distance would be smaller\cite{Costello2025}.

The LungMan phantom is designed to mimic the attenuation of a real human thorax, and we have shown that it has similar phase properties to soft tissue. 
Nevertheless, future experiments using more realistic models, such as large animal thoraces or human cadavers, will be necessary to evaluate diagnostic improvements compared to conventional CT and to study effects not present in our experiments. 
A particularly important effect is the scattering of X-rays by the many air--tissue interfaces created by the lung alveoli. 
This scattering from alveoli is measured in dark-field X-ray imaging \cite{kitchen:2010:xpasrutmpci,schleede:2012:eduxdilcsls,bech:2013:vdpxi,willer:2021:xdcidqepcopddas}, and such scattering has been shown in general to degrade image quality, in particular spatial resolution, in propagation-based imaging \cite{nesterets:2008:odecpi,leatham:2023:xdprofpe,ahlers:2024:xdspi}.

\section{Conclusion}

We have shown that propagation-based X-ray phase contrast is visible in images of an adult human chest phantom, even at the high X-ray energies suitable for such an application. Of the two detectors available at the Australian Synchrotron's Imaging and Medical Beamline, the Eiger photon-counting detector provides the best images. From measurements of imaging quality, we have shown that the optimal energy for adult lung imaging is approximately \qty{70}{\keV}. In addition, we have seen an increase in image quality with propagation distance up to the maximum possible propagation distance within hutch~3B of \qty{7.5}{\m}.
 
\bibliographystyle{abbrv}
\bibliography{references,references-extra}

\section*{Acknowledgements}

J.N.A. acknowledges funding from the Australian Research Training Program (RTP) and an AINSE Ltd. Postgraduate Research Award (PGRA). This project was funded by Research Council (NMHRC) Synergy Grant APP2011204, known as IMPACT. This research was supported by an AINSE Ltd. Early Career Researcher Grant (ECRG).
We thank Patrick Brennan and the University of Sydney, as well as 4DMedical for the loan of Lungman phantoms. 
This research was undertaken on the Imaging and Medical Beamline (IMBL) at the Australian Synchrotron, part of ANSTO. We recognise the scientific and technical assistance of Matthew Cameron.

\section*{Author contributions statement}

J.N.A, M.J.K., K.M.P., K.S.M., and Y.I.N. conceived the experiments. J.N.A, M.D., S.J.A., S.A.H., Y.Y.H., M.K.C., D.H., A.M., C.H., T.E.G., M.J.K., Y.I.N., and K.S.M. conducted the experiments, J.N.A and L.D. processed the data. J.N.A., Y.I.N., and H.B. analysed the results with advice from T.E.G. and J.A.P. Supervision was provided by M.J.K., K.M.P., and K.S.M. The manuscript was prepared by J.N.A. All authors reviewed the manuscript. 

\end{document}


\maketitle

\section{Quality dependence on dose}

\begin{figure}
    \centering
    \includegraphics[scale=1]{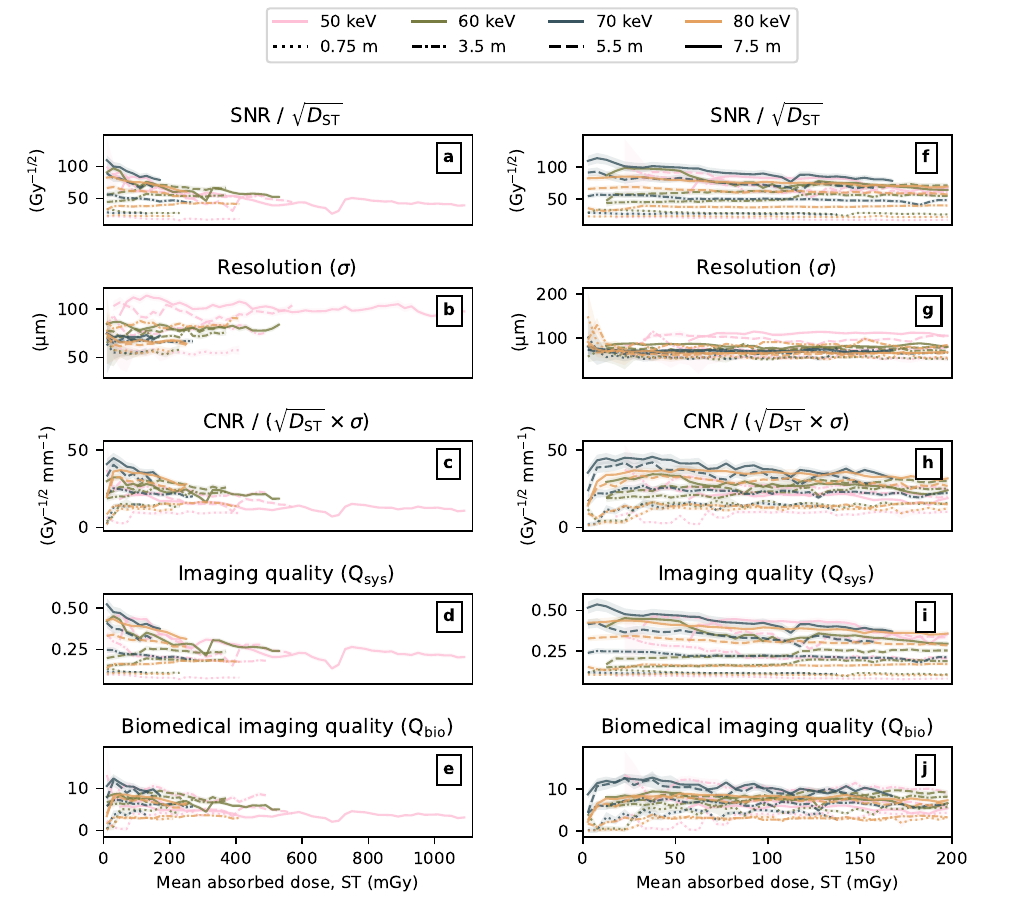}
    \caption{Dose dependence of imaging quality metrics. On the left, the full range of doses is plotted. On the right, the range is restricted to below \qty{200}{\milli\gray}.
    \label{fig:supp-iqa-dose}}
\end{figure}

\begin{figure}
    \centering
    \includegraphics[scale=1]{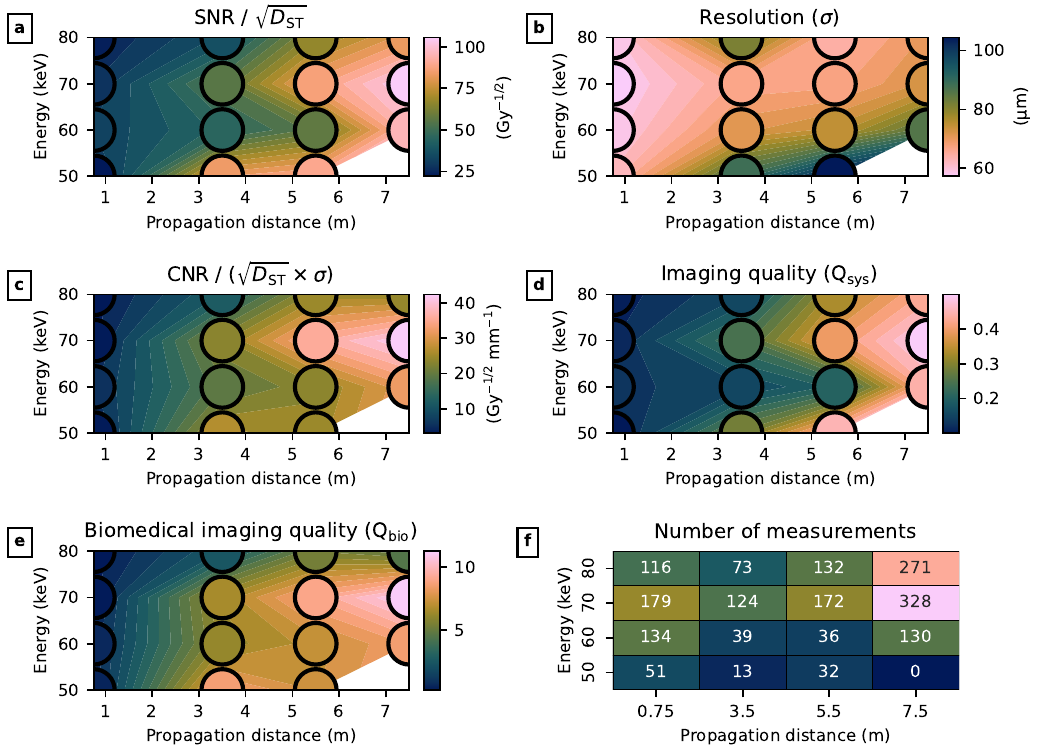}
    {\phantomsubcaption\label{fig:iqaresults-snr}}
    {\phantomsubcaption\label{fig:iqaresults-res}}
    {\phantomsubcaption\label{fig:iqaresults-cnrres}}
    {\phantomsubcaption\label{fig:iqaresults-Qsys}}
    {\phantomsubcaption\label{fig:iqaresults-Qbio}}
    {\phantomsubcaption\label{fig:iqaresults-other}}
    \caption{Results of imaging quality analysis at low dose ($< \qty{50}{\milli\gray}$).
    At \qty{50}{\keV} and \qty{7.5}{\m} there were no slices with a mean absorbed dose below \qty{50}{\milli\gray}, all were at higher dose.
    In the combined imaging metrics, the peak image quality is still at \qty{70}{\keV} and \qty{7.5}{\m}.
    \label{fig:supp-iqa-results}}
\end{figure}

The imaging quality metrics that were used in the main paper included a dose normalisation in order to allow for the comparison of quantities across different CT slices. 
In order to check whether this normalisation correctly removed any dependence on dose, we plotted the metrics against dose in \cref{fig:supp-iqa-dose}.
Overall, the plots are flat with no significant dependence evident.
The imaging quality analysis was repeated on measurements with a mean absorbed dose to soft tissue below \qty{50}{\milli\gray}. 
The results are shown in \cref{fig:supp-iqa-results}.
As before, the best results are achieved at \qty{70}{\keV} and \qty{7.5}{\m}.

